\documentclass[twocolumn,nofootinbib]{revtex4}
\usepackage{graphicx}
\usepackage{latexsym}
\def\be{\begin{equation}}
\def\ee{\end{equation}}
\def\bea{\begin{eqnarray}}
\def\eea{\end{eqnarray}}

\begin{document}
\title{Testing Mass Varying Neutrino With Short Gamma Ray Burst}
\author{Hong Li${}^a$}
\email{hongli@mail.ihep.ac.cn}
\author{Zigao Dai${}^b$}
\author{Xinmin Zhang${}^a$}
\affiliation{${}^a$Institute of High Energy Physics, Chinese
Academy of Sciences, P.O. Box 918-4, Beijing 100049, P. R. China }
\affiliation{${}^b$Department of Astronomy, Nanjing University,
Nanjing 210093, P. R. China }

\begin{abstract}
\hspace{5mm} In this paper we study the possibility of probing for
the absolute neutrino mass and its variation with short Gamma Ray
Burst (GRB). We have calculated the flight time difference between
a massive neutrino and a photon in two different approaches to
mass varying neutrinos. Firstly we parametrize the neutrino mass
as a function of redshift in a model independent way, then we
consider two specific models where the neutrino mass varies during
the evolution of the Quintessence fields. Our calculations show in
general the value of the time delay is changed substantially
relative to a constant neutrino mass. Furthermore our numerical
results show that the flight time delay in these models is
expected to be larger than the duration time of the short GRB,
which opens a possibility of testing the scenario of mass varying
neutrino with the short GRB.
\end{abstract}

\maketitle

In the recent years astronomical observations show that the
Universe is spatially
   accelerating at the present time\cite{z4}. The simplest account
   of this cosmic acceleration seems to be a remnant small
   cosmological constant, but it suffers from the difficulties
associated with the
 fine tuning and coincidence problem, so many
   physicists are attracted with the idea that
the acceleration is driven by dynamical scalar fields, such as
Quintessence. In models of dark energy with a remnant small
cosmological constant or (true or false) vacuum energy $\rho \sim
{( 2 \times 10^{-3} {\rm ev} )}^4$.  This energy scale $\sim
10^{-3}$ ev is smaller than the energy scales in particle physics,
but interestingly is comparable to the neutrino masses. This
indicates a possible connection between the neutrinos and the dark
energy. Furthermore, in Quintessence-like models $m_Q \sim
10^{-33}$ eV, which surprisingly is also connected to the neutrino
masses {\it via} a see-saw formula $m_Q \sim {m_\nu^2 / M_{pl}}$
with $M_{pl}$ the planck mass.

Is there really any connections between the neutrinos and dark
energy? Given the arguments above it is quite interesting to make
such a speculation on this connection. If yes, however in terms of
the language of the particle physics it requires the existence of
new dynamics and new interactions between the neutrino and the
dark energy sector. Recently there are some studies in the
literature on the possible realization of the models on neutrinos
and dark energy\cite{paper30, paper11, paper15, paper31, paper32,
paper33, paper34, paper35, paper36,z24}. One of the interesting
predictions of these models\cite{z25} is that neutrino masses are
not constant, but vary as a function of time during the evolution
of the universe. In this paper we study the possibility of testing
this scenario of mass varying neutrinos with the gamma ray burst.

There are strong evidences for the non-vanishing neutrino masses
from the neutrino oscillation experiments, however the neutrino
masses given by the solar neutrinos and atmosphere neutrinos
experiments are not the absolute values, but mass square
difference: $\Delta m^2_{\bigodot}\sim 8\times 10^{-5} eV^2$(solar
neutrino experiment ), $\Delta m^2_{atm}\sim 2 \times 10^{-3}
eV^2$(atmosphere neutrino experiment)\cite{z17}. The cosmological
observations for example Wilkinson Microwave Anisotropy
Probe(WMAP) and Sloan Digital Sky Survey(SDSS) provide a limit on
the absolute value of the neutrino mass which for a degenerated
spectrum corresponds to $m=0.6$ eV\cite{z18}. Studying the time
delay of neutrinos from the GRB serves another way to measure the
absolute value of the neutrino masses\cite{z28,z16}.

As it is known that most of the energy of a supernova is released
in the form of neutrinos, and therefore it is widely expected that
GRBs are similarly regarded as a high intensity, high energy
neutrino beam with a cosmological baseline which may be detected
in future neutrino telescopes\cite{z1,z2,z3,z20,z21,z28}.

The time delay $t_d$ is defined as the time difference between a
massive neutrino and a photon emitted from a given
source,
\begin{equation}
     t_d\approx \int^{t_0}_t
     a(t^{\prime})dt^{\prime}\frac{1}{2}\frac{m^2}{p^2},
    \end{equation}
     where $p$ is the neutrino energy measured at the detector,
 $m$ is the
neutrino mass, and $a(t)$ is the
    expansion factor of the universe which by normalization is
set to be $a(t_0)=1$ at present time
    $t_0$.
    One can see that the integration over $t$ from the time of
    emission of the photon and neutrino to the present(``$t_0$")
contains information on
    the cosmology and depends on the cosmological parameters.
With the results of the cosmological parameters given by the
recent WMAP and SDSS group,
 $\Omega_{\Lambda_0}=0.73$ and $\Omega_{m_0}=0.27$, we have that:
    \begin{equation}
     t_d\approx
\frac{1}{2}\frac{m^2}{p^2}\int_0^Z\frac{dz^{\prime}}{(1+z^{\prime})^2
     H_0\sqrt{\Omega_{\Lambda_0}+(1+z^{\prime})^3\Omega_{m_0}}}.
    \end{equation}
    In the numerical calculation we will take the Hubble constant
$H_0=1.5359\times
    10^{-42}$ GeV.
In Fig.1 we plot $t_d$ as a function of $p$ the neutrino energy
for four different values of the redshifts. And to get the
possible maximal value of the time delay $t_d$, in the numerical
calculation we have taken the degenerated neutrino mass pattern
and the current cosmological upper limit $m<0.6$eV. By measuring
the $t_d$ one would be able to determine the absolute value of the
neutrino mass, however in general since the photons are trapped
inside the fireball and are released much later, the neutrino and
the photon will not be emitted at the same time. This will add an
systematic error in the measurement of the $t_d$, consequently an
uncertainty in the determination of the neutrino masses. To reduce
this type of systematic error we focus on the short GRB whose
duration is generally less than 2 seconds or even much shorter,
For example Burst And Transient Source Experiment(BATSE) has
discovered a GRB with duration only 5 milliseconds\cite{z19}.
 It has been widely argued that short GRBs are produced by the merger
of two compact objects (e.g., neutron stars or black
holes)\cite{Narayan}, and that their environments are likely to be
low-density media because the merging place is far away from the
birth site of one neutron star. Even so, the predicted afterglows
from short GRBs appear to be detectable with current instruments
in the Swift era\cite{panaitescu}. Once such afterglows are
detected, the redshifts of short GRBs may be measured.

From Fig. 1 one can see that in general the time delay $t_d$
is longer than the duration of the short GRB. For example
for $m=0.6$
eV, $z=2, p=10$Mev, $t_d$ is around 400 seconds;
 especially for the $GRB
030329$
    with redshift $z=0.17$, $t_d$ is 112 seconds.
The time delay for these cases is expected to be detectable in
principle. If not, this will put a limit on the absolute value of
the neutrino mass better than the cosmological limits.

\begin{figure}[htbp]
\includegraphics[scale=0.6]{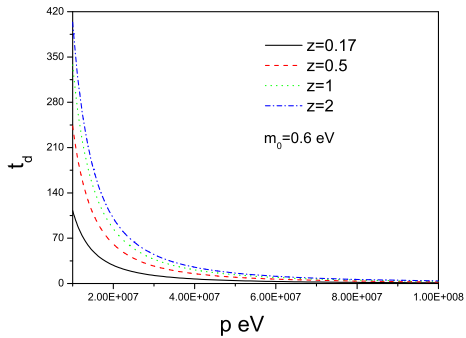}
\caption{$t_d$ (unit in second) as a function of $p$ for different
redshifts z=0.17(solid), 0.5(dashed), 1(dotted), 2(dash dotted).
\label{Fig:spec}}
\end{figure}

In the following we will discuss the possibility of testing on the
variation of the neutrino masses with the short GRB. First of all
we parametrize the variation of the neutrino mass in a model
independent way, then we will take two concrete models of
Quintessence for the calculation of mass variation and the time
delay of the neutrinos. Consider a general case where
  $m_{\nu}$ is an arbitrary
    function of the redshift $Z, ~~{\it i.e.} m(Z)$, one can
    expand it in terms of redshift $z$. For small $z$, we get:
    \begin{eqnarray}\label{double}
   m(Z)&=&m_0+m^{\prime}Z+\frac{1}{2}m^{\prime\prime}Z^2+...
     \nonumber\\
           &=&m_0(1+\frac{m^{\prime}(0)}{m_0}Z+...).
    \end{eqnarray}
    Defining $c\equiv
    \frac{m^{\prime}(0)}{m_0}$, we have
    \begin{equation}
     m=m_0(1+cZ+...),
    \end{equation}
    where $m_0$ is the neutrino mass at $z = 0$.
    Replacing the neutrino mass in (1) by (5) we obtain that
    \begin{equation}
      t_d\approx \frac{1}{2}\frac{m_0^2}{p^2}
   \int_0^Z\frac{(1+cz^{\prime})^2dz^{\prime}}{(1+z^{\prime})^2
     H_0\sqrt{\Omega_{\Lambda_0}+(1+z^{\prime})^3\Omega_{m_0}}}.
    \end{equation}

In Fig. 2 we plot $t_d$ as a function of $\frac{m_0}{p}$.
Note that here the neutrino mass in (5) varies as a function of
redshift $z$.
    \begin{figure}[htbp]
\includegraphics[scale=0.6]{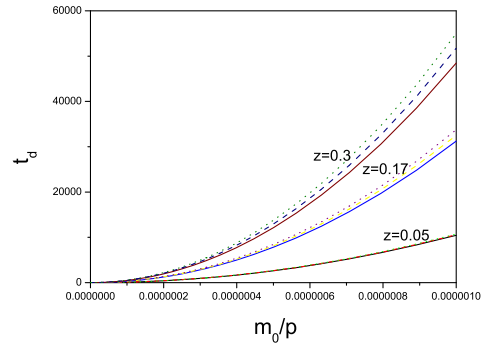}
\caption{ $t_d$ (unit in second) as a function of $\frac{m_0}{p}$
for different coefficient $c$: c=0(solid), 0.5(dashed), 1(dotted).
\label{Fig:spec}}
\end{figure}
From Fig. 2 one can see that
the variation of the neutrino mass does change the value of $t_d$
compared to that if the neutrino mass is constant ($c= 0$).

The neutrino mass in the standard model of electroweak theory
comes from a dimension five operator
     \begin{equation}
     {\cal L}=\frac{2}{f}\nu\nu\phi\phi+h.c.  ,
    \end{equation}
and  $m_{\nu}\sim\frac{v^2}{f}$ where
    $f$ is the scale of the new physics generating the neutrino mass and
$v$ is
    the vacuum expectation value of the Higgs fields.
In order to have a mass varying neutrino we introduce a coupling
of Quintessence to the
     neutrinos,
     \begin{equation}
     \beta\frac{Q}{M_{pl}}\frac{2}{f}\nu\nu\phi\phi+h.c. ,
    \end{equation}
    where $\beta$ is the coefficient which characterizes the
    strength of Quintessence interacting with the neutrinos and
    generally one requires $\beta<4\pi$ to make the effective
    lagrangian description reliable.

Combining (8) with (7)
    the neutrino mass is given by:
    \begin{equation}
     m=m_0\frac{1}{1+\beta\frac{Q_0}{M_{pl}}}(1+\beta\frac{Q}{M_{pl}}) ,
    \end{equation}
    where $m_0$ and $Q_0$ are the
    neutrino mass and the value of Quintessence at
present
    time, and $M_{pl}$ is the Plank scale.

    Now the formula for the time-delay is given by
     \begin{equation}
    t_d=\frac{1}{2}(\frac{m_0}{p})^2\frac{1}{(1+\beta\frac{Q_0}{M_{pl}})^2}
    \int^{t_0}_t(1+\beta\frac{Q}{M_{pl}})^2a(t^{\prime})dt^{\prime}  .
    \end{equation}

    To evaluate $t_d$ we need to know the evolution of the
Quintessence which can be obtained by solving the following
    equations of motion of the Quintessence. For a flat
    Universe they are,
    \begin{equation}
    H^2=\frac{8\pi G}{3}(\rho_B+\frac{\dot{Q}^2}{2}+V(Q)),
    \end{equation}
    \begin{equation}
    \ddot{Q}+3H\dot{Q}+V^{\prime}(Q)=0,
    \end{equation}
    \begin{equation}
    \dot{H}=-4\pi G((1+\omega_B)\rho_B+\dot{Q}^2),
    \end{equation}
    where $\rho_B$ and $\omega_B$ represent the energy density and
    the equation-of-state of the background fluid respectively,
    for example $\omega_B=\frac{1}{3}$ in radiation-dominated and
    $\omega_B=0$ in the matter-dominated Universe.

    For a numerical study, we consider a model of Quintessence
    with a inverse power-law potential\cite{z22},
    \begin{equation}
    V=V_0Q^{-\alpha}.
    \end{equation}
    This type of model is shown\cite{z22,z23} to have the property of
    tracking
    behavior. In the calculation we take $\alpha=0.5$ which gives rise
to  $\omega_{Q_0}<-0.78$ constrained by the WMAP.
     In Fig. 3 we plot the numerical value of $t_d$ .
    \begin{figure}[htbp]
\includegraphics[scale=0.6]{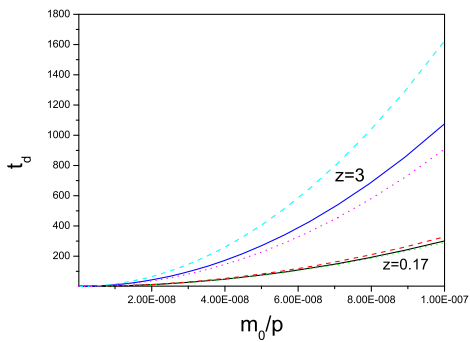}
\caption{Plot of $t_d$ (unit in second) as a function of
$\frac{m_0}{p} $. The three curves correspond to parameter
$\beta$=0(solid), -3(dash), 3(dot) respectively for a given
redshift. \label{fig:spec}}
\end{figure}
From Fig. 3 one can see the time delay $t_d$ is enhanced
substantially relative to a constant value of the neutrino mass.
For example, if $(\frac{m_0}{p})^2\sim 10^{-7}$, and $\beta=-3$,
$t_d$ can reach 1624 seconds at $ z = 3$. For comparison a
constant neutrino mass $\beta = 0$ reduces $t_d$ to 1075 seconds.

For the case of hierarchical neutrino mass patter, the heavy
neutrino mass set by the atmosphare neutrino oscillation is $m
\sim 0.05$eV. With this value one can see from Fig.3 that $t_d$ is
only $0.5$
 seconds for the constant-mass neutrinos with energy $p=12$ Mev and redshift
 $z=3$.
This value is much smaller than the short GRB duration and will
be difficult to be observable.
 If the neutrino mass varies, as shown in Fig. 3
$t_d$ will be enhanced substantially. Numerically Fig.3 shows that
$t_d$ can be increased to 3 seconds.

Certainly the results shown in Fig.3 depends on the
Quintessence model. For an illustration , we consider another
Quintessence model
\begin{equation}
    V=V_0exp(\frac{\lambda}{Q}).
    \end{equation}
    In this calculation we take $\lambda=0.5 M_{pl}$.
From Fig.4 one can see that different model of Quintessence does
give a different result of $t_d$, but the $t_d$ is in general larger
than the duration time of the short GRB.

\begin{figure}[htbp]
\includegraphics[scale=0.6]{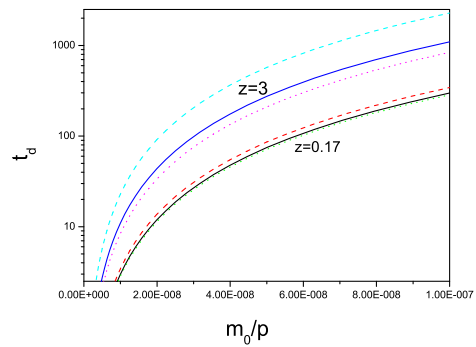}
\caption{Plot of $t_d$ (unit in second) as a function of
$\frac{m_0}{p}$. The three curves correspond to parameters
$\beta$=0(solid), -2(dash), 10(dot) respectively for a given
redshift. \label{fig:spec}}
\end{figure}

In summary we have in this paper discussed the possibility of using the
Short GRB
to probe for the absolute neutrino
masses and its variation. By a detailed calculation we have shown that
with the current cosmological limits on the degenerated neutrino masses
the time delay $t_d$ will be in general longer than the time of the
duration,
furthermore the changes to the $t_d$ caused by the variation of the
neutrino masses are also expected to be larger than the time of the
duration.
 Our
results indicate the possibility of testing the scenario of mass varying
neutrino in the future neutrino telescope.

{\bf{Acknowledgments:}}We are indebted to Drs. Minzhe Li, Bo Feng
and Xiaojun Bi for enlightening discussions and Profs. Roberto
Peccei, Sandip Pakvasa and Dr. Heinrich Paes for reading the
manuscript. We thank Dr. Nan Liang for helpful discussions on
GRBs. This work was supported in part by National Natural Science
Foundation of China (grant Nos. 90303004, 19925523, 10233010 and
10221001) and by Ministry of Science and Technology of China(
under Grant No. NKBRSF G19990754).

\end{document}